\documentclass[prb,twocolumn,showpacs]{revtex4}

\usepackage{graphicx}
\begin{document}

\title{
   Pair-distribution functions of correlated composite fermions}

\author{
   Arkadiusz W\'ojs$^{1,2}$, 
   Daniel Wodzi\'nski$^1$, and 
   John J. Quinn$^2$}

\affiliation{
   $^1$Wroclaw University of Technology, 50-370 Wroclaw, Poland\\
   $^2$University of Tennessee, Knoxville, Tennessee 37996, USA}

\begin{abstract}
   Pair-distribution functions $g(r)$ of Laughlin quasielectrons
   (composite fermions in their second Landau level) are calculated 
   in the fractional quantum Hall states at electron filling factors 
   $\nu_e=4/11$ and 3/8.
   A shoulder in $g(r)$ is found, supporting the idea of cluster 
   formation.
   The intra- and inter-cluster contributions to $g(r)$ are identified, 
   largely independent of $\nu_e$.
   The average cluster sizes are estimated; pairs and triplets of 
   quasielectrons are suggested at $\nu_e=4/11$ and 3/8, respectively.
\end{abstract}
\maketitle

\section{Introduction}

Pan et al.\cite{pan03} have recently observed the fractional quantum Hall 
(FQH) effect\cite{tsui82,laughlin83} in a spin-polarized two-dimensional 
electron gas (2DEG) at the $\nu_e={4\over11}$, ${3\over8}$, and ${5\over13}$ 
fillings of the lowest Landau level (LL).
In the composite fermion (CF) model,\cite{jain89,lopez91} these values 
correspond to the fractional fillings $\nu={1\over3}$, ${1\over2}$, 
and ${2\over3}$ of the second CF LL, respectively.
In Haldane's hierarchy picture\cite{haldane83} of these states, Laughlin 
quasielectrons (QE's) fill (the same) fraction $\nu$ of their LL.
The most striking conclusion from Pan's discovery is that the CF's (or 
QE's) can also form incompressible states when partially filling a LL.
This could not be predicted by a simple analogy with known fractional 
electron liquids (Laughlin,\cite{laughlin83} Jain,\cite{jain89} or 
Moore--Read\cite{moore91} states), because of a different form of 
QE--QE interaction,\cite{sitko96,hierarchy,lee01} therefore yielding 
qualitatively different QE--QE correlations.

Although several numerical studies of interacting QE's have been 
reported\cite{lee01,mandal02,qeclust,chang04} and ideas such as CF 
flavor-mixing,\cite{peterson04} QE pairing,\cite{flohr03,qepair} or 
stripes\cite{shibata05} were invoked, the correlations responsible for 
the FQHE at $\nu_e={4\over11}$ and ${3\over8}$ are not yet understood.
It has not even been settled if these FQH states are isotropic,
and the energies of liquid and solid phases were compared recently
\cite{goerbig04} (although the Laughlin form was arbitrarily 
assumed for the liquid).

Sometimes overlooked is a general connection\cite{haldane85,parentage} 
between the form of Haldane pseudopotential,\cite{haldane87} occurrence 
of Laughlin correlations, and the validity of CF transformation.
Actually, the form of QE--QE interaction is known from independent 
calculations,\cite{sitko96,hierarchy,lee01} and Laughlin correlations 
among the QE's have been ruled out using both a general pseudopotential 
argument\cite{hierarchy} and a direct analysis of many-QE wavefunctions.
\cite{qeclust} 
In this paper we refer to the following well-established facts:

(i) The QE--QE Haldane pseudopotential\cite{haldane87} is known from 
exact diagonalization of the Coulomb interaction among electrons in the 
lowest LL.\cite{sitko96,hierarchy,lee01}
Since there are no unchecked assumptions in such a calculation, it must 
be regarded a ``numerical experiment.''
Neither finite-size errors, lowest-LL restriction, finite 2DEG width, 
nor other details of realistic experimental systems affect the dominant 
feature of the pseudopotential which is the {\em lack of strong QE--QE 
repulsion at short range}.

(ii) The QE's {\em do not}\cite{hierarchy,qeclust} have Laughlin 
correlations at $\nu={1\over3}$ corresponding to $\nu_e={4\over11}$.
The Moore--Read half-filled state {\em is not}\cite{qeclust,3body} an 
adequate description of QE--QE correlations at $\nu={1\over2}$ 
corresponding to $\nu_e={3\over8}$.

(iii) A sequence of nondegenerate finite-size QE ground states with 
a gap, extrapolating to $\nu={1\over3}$ has been found\cite{qeclust} 
on a sphere. 
Although spherical geometry is not adequate for studying crystal or other 
broken-symmetry phases, the identified states appear incompressible and 
have the lowest energy of all QE liquids (considerably below the Laughlin 
state).

To address the problem of correlations at $\nu_e={4\over11}$, ${3\over8}$, 
and ${5\over13}$ we calculate pair-distribution functions $g(r)$ in the 
incompressible liquid ground states of up to $N=14$ QE's.
Their comparison with the (known) curves of the Laughlin and Moore--Read 
states implies a different nature of the QE correlations in these novel 
FQH states.
It shows that their incompressibility cannot be explained by a simple 
analogy between the QE and electron liquids, and suggests that different 
wavefunctions need be proposed for the correlated CF's.
Unfortunately, the calculated $g(r)$ are of little help in a precise 
definition of these wavefunctions, even though some qualitative 
statements can be made about the QE correlations. 

From our finite-size results we identify and analyze the size-independent 
features in $g(r)$: the $\sim r^2$ behavior at short range and a shoulder 
at a medium range, and argue that they are consistent with the 
idea\cite{qeclust} of QE cluster formation.
Short- and long-range contributions to $g(r)$ are found, describing
correlations between the QE's from the same or different clusters.
Both intra- and inter-cluster QE--QE correlations depend rather weakly 
on $\nu$.
The average size of the clusters is estimated; it seems that the 
QE's form pairs at $\nu={1\over3}$ and triplets at $\nu={1\over2}$.
A similar analysis of $g(r)$ carried out for the Moore--Read state 
reveals a qualitatively different behavior.

\section{Model}

\subsection{Haldane sphere}

The numerical calculations have been carried out in Haldane's spherical 
geometry,\cite{haldane83} convenient for the exact study of short-range 
correlations.
In this model, the lowest LL for particles of charge $q$ is 
a degenerate shell of angular momentum $l=Q$.
Here $2Q$ is the strength of Dirac monopole in the center of the sphere 
defined in the units of elementary flux $\phi_0=hc/q$ as $2Q\phi_0=
4\pi R^2 B$, the total flux of the magnetic field $B$ through the 
surface of radius $R$.
Using the usual definition of the magnetic length, $\lambda=
\sqrt{\hbar c/qB}$, this can be written as $l\lambda^2=R^2$.
In the following, $\lambda$ denotes the QE magnetic length 
corresponding to the fractional charge $q=-e/3$.

The relative ($\mathcal{R}$) and total ($L$) pair angular momenta are 
related via $L=2l-\mathcal{R}$.
For fermions, $\mathcal{R}$ is an odd integer, and it increases with
increasing average pair separation $\sqrt{\left<r^2\right>}$.
The interaction (within the lowest LL) is entirely determined by Haldane 
pseudopotential defined as the pair interaction energy $V$ as a function 
of $\mathcal{R}$.

\subsection{Exact diagonalization}

Recently, we have identified\cite{qeclust} the series of finite-size 
spin-polarized states that in the thermodynamic limit describe the 
FQHE at $\nu_e={4\over11}$ and ${3\over8}$.
To do so, we have carried out extensive exact-diagonalization 
calculations for interacting QE's (particles in the second CF LL).
On Haldane sphere, $N$ fermionic QE's were confined in a standard 
way to an angular momentum shell of degeneracy $\Gamma=2l+1$, 
corresponding to the QE filling factor $\nu\sim N/\Gamma$, and
the Haldane QE--QE pseudopotential $V(\mathcal{R}$) was taken from 
earlier calculations.\cite{sitko96,hierarchy,lee01}  

Regardless of the electron layer width $w$, magnetic field $B$,
or other experimental parameters, the dominant feature of 
$V(\mathcal{R})$ is strong repulsion at $\mathcal{R}=3$.
This feature alone determines the wavefunctions at ${1\over3}
\le\nu\le{1\over2}$ (with the QE--QE correlations consisting 
of maximum possible avoidance of Haldane pair amplitude $\mathcal{G}$
at $\mathcal{R}=3$), which are hence virtually insensitive to the 
(sample-dependent) details of $V(\mathcal{R})$.
This justifies model calculations using $V(\mathcal{R})$ of 
Refs.~\onlinecite{sitko96,hierarchy,lee01}.
Actually, a model pseudopotential as simple as 
$V=\delta_{\mathcal{R},3}$ is sufficient to reproduce correct
correlations and incompressibility at $\nu_e={4\over11}$ 
or ${3\over8}$.

\section{Numerical Results}

\subsection{Energy spectra}

\begin{figure}
\resizebox{3.4in}{1.7in}{\includegraphics{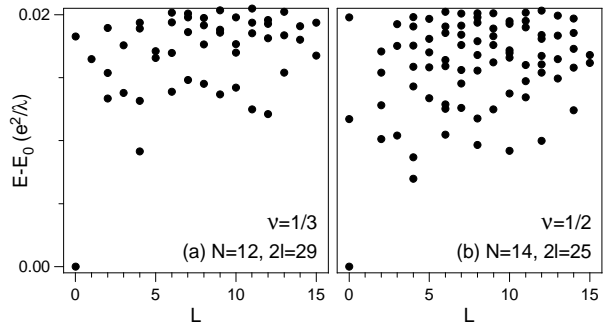}}
\caption{
   Excitation energy spectra (energy $E$ as a function of total 
   angular momentum $L$; $E_0$ is the ground state energy)
   of $N$ interacting QE's on a sphere, at the values of CF LL 
   degeneracy $\Gamma=2l+1$ corresponding to the incompressible 
   ground states at the QE filling factors $\nu={1\over3}$ (a) 
   and ${1\over2}$ (b).}
\label{fig1}
\end{figure}

The numerical results carried out for $N\le14$ (two sample spectra 
are displayed in Fig.~\ref{fig1}) showed\cite{qeclust} a sequence 
of nondegenerate (i.e., at the total angular momentum $L=0$) ground 
states at $2l=N/\nu-\gamma$ with $\nu={1\over3}$ and $\gamma=7$.
The significant and well-behaved (as a function of $N$) excitation gap 
along this sequence strongly suggests that it represents the infinite 
$\nu_e={4\over11}$ FQH state observed in experiment.\cite{pan03}
The value $\gamma\ne3$ precludes Laughlin correlations among QE's 
in this state (earlier ruled out indirectly, based on the form of 
QE--QE pseudopotential\cite{hierarchy}), i.e., an idea that the
$\nu_e={4\over11}$ state is simply a Haldane hierarchy state of
Laughlin-correlated CF's.
While the exact correlations in this (known only numerically
for a few consecutive $N$) ground state have not yet been
defined, their vanishing degeneracy ($L=0$) implies that
they describe a QE liquid, rather than a broken-symmetry state 
(such as liquid-crystal nematic states proposed\cite{musaelian96} 
in the context of FQHE at different values of $\nu$).

Another sequence was anticipated at $2l=2N-\gamma$ to represent
the infinite $\nu_e={3\over8}$ FQH state.
However, the only ground state with a significant gap and remaining 
outside of the $\nu={1\over3}$ sequence (or its particle-hole symmetric 
$\nu={2\over3}$ sequence at $2l={3\over2}N+2$) occurs\cite{qeclust} 
for $N=14$ and $2l=25$ (and it also has $L=0$).
These values of $(N,2l)$ happen to belong to a $2l=2N-3$ series
representing the Moore--Read (pfaffian) paired state, but the overlap 
between the two turns out nearly zero.\cite{qeclust,3body}
Moreover, the ground states for the two neighboring even (as 
appropriate for a hypothetically paired state) values of 
$N=12$ and 16 (and $2l=21$ and 29) have $L>0$ and no gap, the 
value of $2l=17$ for $N=10$ coincides with the $\nu={2\over3}$
sequence (so that only for $N>8$ can the filling factor $\nu$ 
be meaningfully assigned), and we are unable to compute the spectra 
for $N\ge18$.
Nevertheless, despite little evidence available from numerical
diagonalization, the ground state for $N=14$ and $2l=25$ (and 
its particle-hole counterpart at $N=12$ and the same $2l=25$)
may possibly represent the $\nu_e={3\over8}$ FQH state (i.e., 
have similar correlations causing incompressibility).

\subsection{Pair-distribution functions}

\begin{figure}
\resizebox{3.4in}{1.7in}{\includegraphics{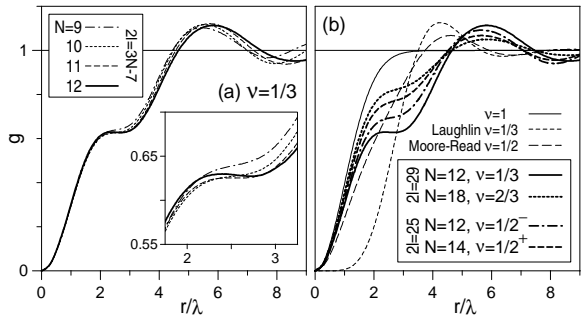}}
\caption{
   QE--QE pair-distribution functions $g(r)$ of the incompressible
   ground states at different QE filling factors $\nu$.
   (a) curves for $\nu={1\over3}$ and different QE numbers $N$;
   (b) curves for QE's at different $\nu$ (thick lines) compared 
   to some known incompressible states of electrons.}
\label{fig2}
\end{figure}

The QE--QE pair-distribution functions $g(r)$ have been calculated for 
the incompressible many-QE ground states as expectation values of the 
appropriate pair interaction, 
\begin{equation}
   g(r)=(2/N)^2\left<\delta(R\theta-r)\right>.
\label{eq1}
\end{equation}
Here, $\theta$ is the relative angle on a sphere, so that $r$ measures 
interparticle distance along the surface (rather than chord distance).
More accurately, $r$ is the distance between the centers of extended
QE's (note that in the calculation of many-QE wavefunctions, the system 
of QE's is mapped onto the lowest LL of point charges interacting
through an effective pseudopotential).
The prefactor in Eq.~(\ref{eq1}) ensures proper normalization, 
$g(\infty)\rightarrow1$.
Denoting infinitesimal area by $dS=2\pi R^2\,d(\cos\theta)$ or 
(in magnetic units) by $ds=dS/2\pi\lambda^2$, we get an equivalent
normalization condition, 
\begin{equation}
   \int[1-g(r)]\,ds={2l\over N}\rightarrow\nu^{-1}
\label{eq2}
\end{equation}
in large systems.
Since $ds=l\,d(\cos\theta)$, a ``local filling factor'' can also be 
defined as $\nu(r)=dN/ds=(N/2l)\,g(r)$, and it satisfies $\nu(\infty)=
\nu$ and $\int\nu(r)ds=N-1$.

The results for the $\nu={1\over3}$ sequence at $2l=3N-7$ are shown in 
Fig.~\ref{fig2}(a).
Similarity of all four curves is evident, indicating size-independent
form of correlations (hence, describing an infinite system), with a 
well-developed shoulder around $r\approx2.5\lambda$.
Similar shoulders occur in $g(r)$ of all incompressible ground states 
at $\nu={2\over3}$ or ${1\over2}$ (the $\nu={2\over3}$ sequence at $2l
={3\over2}N+2$ is obtained from $2l=3N-7$ by replacing $N$ with $\Gamma
-N$, while at $\nu={1\over2}$ there are two particle--hole conjugate 
sequences at $2l=2N-3$ and $2N+1$, denoted by $\nu={1\over2}^\pm$).
The four curves representative of $\nu={1\over3}$, ${2\over3}$, and 
${1\over2}^\pm$ are shown in Fig.~\ref{fig2}(b).
They are all clearly different from those marked with thin lines and 
describing correlations known for other incompressible FQH states (full 
LL, Laughlin $\nu={1\over3}$ state, or Moore--Read half-filled state).
This is a direct indication of a different nature of QE--QE correlations
responsible for the FQHE at $\nu_e={4\over11}$ and ${3\over8}$.

Let us stress that although the QE--QE interactions are not known 
with great accuracy, the correlation functions in Fig.~\ref{fig2}
are rather insensitive to the details of $V(\mathcal{R})$, as long
as the dominant repulsion occurs at $\mathcal{R}=3$ (which seems
to be universally true in the systems studied experimentally).
This insensitivity is reminiscent of the Laughlin wavefunction, 
which also very accurately describes the actual $\nu={1\over3}$ 
ground state for a wide class of electron--electron pseudopotentials.
However, while the avoidance of $\mathcal{R}=1$ by the electrons
in the lowest LL can be elegantly described by flux attachment
in the CF picture, no similar model has been proposed yet for the 
avoidance of $\mathcal{R}=3$ by the QE's.
Therefore, knowing the $g(r)$ curves of QE's and understanding their 
correlations, we still cannot write their wavefunctions.

\subsection{Gaussian deconvolution}

\begin{figure}
\resizebox{3.4in}{1.7in}{\includegraphics{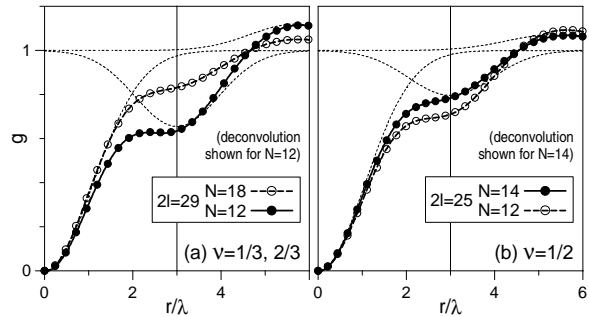}}
\caption{
   Gaussian deconvolution of the QE--QE pair distribution functions $g(r)$:
   dots -- data of Fig.~\ref{fig2}(b);
   lines -- fits.}
\label{fig3}
\end{figure}

\begin{table}
\caption{
   Gaussian deconvolution parameters for QE--QE pair-distribution
   functions shown in Figs.~\ref{fig2}(b) and \ref{fig3}.}
\label{tab1}
\begin{ruledtabular}
\footnotesize
\begin{tabular}{l|ccc|ccc|ccc}
$\phantom{a}\nu$ & $A_0$&$\delta_0$&$\sigma_0$ 
& $A_1$&$\delta_1$&$\sigma_1$ 
& $A_2$&$\delta_2$&$\sigma_2$ \\
\hline
$1/3$   & 1&0&1.0989 & 0.3450&3&0.9412 & -0.1199&5.6905&1.0298 \\
$2/3$   & 1&0&1.0419 & 0.1535&3&0.9361 & -0.0530&5.6655&0.9987 \\
$1/2^+$ & 1&0&1.0626 & 0.2034&3&0.9475 & -0.0741&5.4041&1.1011 \\
$1/2^-$ & 1&0&1.0896 & 0.2755&3&0.9431 & -0.1005&5.4156&1.0903 
\end{tabular}
\end{ruledtabular}
\end{table}

The curves of Fig.~\ref{fig2}(b) can be accurately deconvoluted using 
gaussians, $G(r/\lambda)=A\exp[-(r/\lambda-\delta)^2/2\sigma^2]$.
This is shown in Fig.~\ref{fig3} where the symbols mark the exact 
data of Fig.~\ref{fig2}(b) and the lines give the (nearly perfect) 
fits using three gaussians, $g=1-G_0-G_1-G_2$ (sufficient for 
$r\le6\lambda$).
The fitted values of $[A_i,\delta_i,\sigma_i]$ for all four curves 
are listed in Tab.~\ref{tab1}.
Note that $A_0=1$, $\delta_0=0$, and $\delta_1=3$ for all curves
(the latter value being least obvious, but probably resulting from 
the avoidance of the same $\mathcal{R}_3=3$ by the QE's at all values 
of $\nu$).
The values of the $G_2$ parameters are not very meaningful when the 
next term in the approximation ($G_3$) is neglected.
The clearest difference between the four curves is in $A_1$.

\subsection{Short/long-range deconvolution}

\begin{figure}
\resizebox{3.4in}{1.7in}{\includegraphics{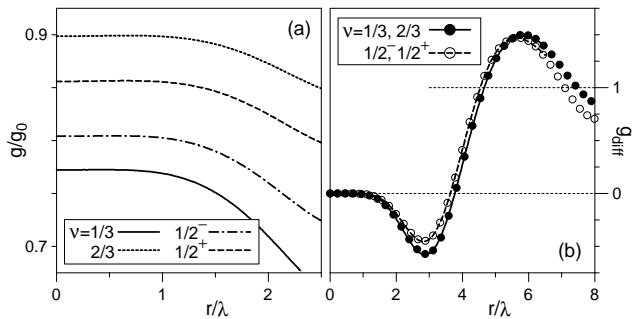}}
\caption{
   (a) Ratio of QE--QE pair-distribution functions $g(r)$ to 
   $g_0(r)$ of a full lowest LL for different incompressible
   QE ground states;
   (b) the ``remainder'' $g_{\rm diff}(r)$ defined by Eq.~(\ref{eq3}).}
\label{fig4}
\end{figure}

It appears more physically meaningful to decompose $g(r)$ into $g_0=
1-\exp(-r^2/2\lambda^2)$ describing a full lowest LL\cite{jancovici81} 
and a (properly normalized) ``remainder'' $g_{\rm diff}$,
\begin{equation}
   g(r)=\alpha\,g_0(r)+(1-\alpha)\,g_{\rm diff}(r).
\label{eq3}
\end{equation}
For each $g(r)$, parameter $\alpha$ is calculated as the limit of 
$g/g_0$ at $r\rightarrow0$.
It is clear from Fig.~\ref{fig4}(a) that $g(r)$ is accurately 
approximated by $\alpha\,g_0(r)$ within a finite area or a radius 
$\sim\lambda$ for all four ground states of Fig.~\ref{fig2}(b).
The numerical values of $\alpha$ are 0.772, 0.804, 0.856, and 
0.899 for $\nu={1\over3}$, ${1\over2}^-$, ${1\over2}^+$, and 
${2\over3}$, respectively. 
Evidently, $\alpha$ is size-dependent (e.g., the pair of values for
$\nu={1\over2}^\pm$ must converge to the same thermodynamic limit).

The four curves $g_{\rm diff}(r)$ calculated from Eq.~(\ref{eq3}) are 
plotted in Fig.~\ref{fig4}(b). 
Symbols and lines mark the exact data and the three-gaussian fits 
of Tab.~\ref{tab1}, respectively.
We note that:
(i) For the pairs of particle--hole conjugate states ($N=12$, 18 at 
$2l=29$ and $N=12$, 14 at $2l=25$), the $g_{\rm diff}(r)$ curves are 
{\em identical}.
(ii) The curves obtained for $\nu={1\over3}$ and ${1\over2}$ are very 
similar (and possibly identical in large systems); they all vanish at 
short range and have a minimum at $r\approx3\lambda$ and a maximum at 
$r\approx5.5\lambda$.

\section{Discussion}

\subsection{QE clustering}

Some information about the form of QE--QE correlations can be easily 
deduced from the form of interaction pseudopotential $V(\mathcal{R})$, 
which is simply the interaction hamiltonian defined only for those 
pair states allowed in the lowest LL.
In low-energy many-body states the particles generally tend to avoid 
pair eigenstates with high interaction energy, which means minimization 
of the corresponding Haldane pair amplitude $\mathcal{G}$.
If repulsion $V$ decreases sufficiently quickly\cite{haldane85} as a 
function of $\mathcal{R}$ (the exact criterion being\cite{parentage} 
that $V$ decreases sublinearly as a function of $\sqrt{\left<r^2\right>}$), 
the smallest value of $\mathcal{R}=1$ is avoided.
This Laughlin type of correlation is elegantly described by attachment 
of $2p=2$ fluxes to each particle in the CF transformation.
In a Laughlin-correlated state, each particle avoids being close to 
any other particle (as much as possible at a given finite $\nu$).

When short-range repulsion weakens ($V$ at $\mathcal{R}=1$ decreases 
compared to $V$ at $\mathcal{R}\ge3$), Laughlin correlations disappear
and can be replaced by pairing or formation larger clusters. 
Pairs\cite{flohr03,qepair} or clustering\cite{qeclust} were suggested 
by several authors for the QE's.
This idea was justified by an observation that QE--QE pseudopotential 
nearly vanishes at $\mathcal{R}=1$ and is strongly repulsive at 
$\mathcal{R}=3$, causing an increase of $\mathcal{G}(1)$ and a 
simultaneous is decrease of $\mathcal{G}(3)$ compared to the 
Laughlin-correlated state.\cite{qeclust} 

The assumption that QE's form clusters naturally explains a shoulder
in $g(r)$, and allows one to interpret $g_0$ and $g_{\rm diff}$ as the 
intra- and inter-cluster QE--QE correlations, i.e. the short- and 
long-range contributions to $g$, corresponding to the QE pairs 
belonging to the same or different clusters, respectively.
The vanishing of $g_{\rm diff}(r)$ at short-range reflects isolation 
of QE's belonging to different clusters.
The reason why $g_{\rm diff}$ is not positive definite is that 
intra-cluster correlations are accurately described by $g_0$ only 
within a certain radius.
In other words, the actual inter-cluster contribution to $g$ is not 
{\em exactly} given by $g_{\rm diff}$ defined by Eq.~(\ref{eq3}).
Nevertheless, the following two conclusions remain valid:
(i) the intra- and inter-cluster QE--QE correlations are similar at 
$\nu={1\over3}$, ${1\over2}$, and ${2\over3}$, with the respective 
correlation-hole radii $\varrho_0\sim\lambda$ and $\varrho_1\sim4
\lambda$; and (ii) the cluster size $K$ depends on $\nu$.

Similar form of $g(r)$ was found\cite{musaelian96} for broken-symmetry 
Laughlin states, in which the shoulder results from angular averaging 
of an anisotropic function $g(r,\phi)\sim r^2$ or $r^6$, depending on 
$\phi$.
However, the present case of QE's is different, because $g(r)$ is 
isotropic (wavefunctions have $L=0$) and the shoulders result from 
{\em radial} averaging of inter- and intra-cluster correlations,
(beginning as $\sim r^2$ and a higher power of $r$ at short range, 
respectively).

\subsection{Average cluster size}

In a clustered state, the (average) cluster size $K$ is connected 
to $\alpha$, and the form of $g_{\rm diff}$ depends on correlations 
between the clusters.
The values of $K$ at $\nu={1\over3}$ or ${1\over2}$ can be estimated 
by comparison of the actual parameters $\alpha$ with those predicted 
for the hypothetical states of $N$ particles arranged into $N/K$ 
{\em independent} $K$-clusters.
By independence of the clusters we mean that inter-cluster correlations
do not affect the local filling factor $\nu(r)$ at short range.
For a single cluster, which on a sphere is the $K$-particle state 
with the maximum total angular momentum $L=Kl-{1\over2}K(K-1)$, the 
$\nu_K(r)$ depends on the surface curvature and thus (through 
$R/\lambda=\sqrt{l}$) on $2l$.

\begin{table}
\caption{
   Parameters $\beta_K$ of the short-range approximation
   $\nu(r)\sim\beta\,g_0(r)$ obtained for independent
   clusters of size $K$.}
\label{tab2}
\begin{ruledtabular}
\footnotesize
\begin{tabular}{c|ccccc}
$2l$ & $\beta_2$ & $\beta_3$ & $\beta_4$ & $\beta_5$ & $\beta_6$ \\ 
\hline
25       & 0.2768 & 0.4196 & 0.5110 & 0.5765 & 0.6269 \\
29       & 0.2730 & 0.4134 & 0.5029 & 0.5669 & 0.6159 \\
60       & 0.2609 & 0.3938 & 0.4778 & 0.5372 & 0.5821 \\
$\infty$ & 0.2500 & 0.3763 & 0.4555 & 0.5110 & 0.5527
\end{tabular}
\end{ruledtabular}
\end{table}

We have calculated the prefactors $\beta_K$ of the short-range 
approximation $\nu_K(r)\approx\beta_K g_0(r)$ for different values 
of $K$ and $2l$ and listed some in Tab.~\ref{tab2} (note that 
$\nu_2(r)$ is known exactly).
These coefficients are to be compared with $\beta=(N/2l)\,\alpha$ 
of the incompressible $N$-QE states obtained from diagonalization.
Of course, this approach is somewhat questionable as one generally 
cannot deduce the precise cluster size from the short-range behavior 
of $g(r)$ for the following reasons:
(i) $K$ is not a well-defined (conserved) quantum number;
(ii) $\nu={1\over3}$ states occur for all $N$ (not only those divisible
by two or three) which means that all clusters cannot have the same $K$;
(iii) parameters $\alpha$ and $\beta$ are size-dependent and their
extrapolation to large systems is not very reliable based on limited
number of $N$-QE systems we are able to diagonalize;
(iv) inter-cluster exchange of QE's makes the ``independent-cluster''
picture only an approximation.

Fortunately, we can use the Moore--Read states (known to be paired
\cite{moore91,3body}) as a test.
Our calculation (for details see Sec.~\ref{sec_pff}) for $N=14$ and 
$2l=25$ gives $\beta_{\rm MR}=0.336$, somewhat larger than $\beta_2$.
Hence, we shall assume that $\beta_K$ in general underestimates the 
actual value of $\beta$ in a many-body $K$-clustered state.

For the QE's, we got $\beta=0.319\approx\beta_{\rm MR}$ for $N=12$
and $2l=29$ ($\nu={1\over3}$), and $\beta=0.479$ for $N=14$ and $2l=25$ 
($\nu={1\over2}^+$; directly comparable with the Moore--Read state).
With appropriate reservation, we can hence risk a hypothesis that
QE's (on the average) form pairs at $\nu={1\over3}$ and triplets at 
$\nu={1\over2}$ (possible triplet formation might turn out especially 
intriguing in the context of parafermion statistics\cite{read}).

\subsection{Comparison with Moore--Read state}
\label{sec_pff}

\begin{figure}
\resizebox{3.4in}{1.7in}{\includegraphics{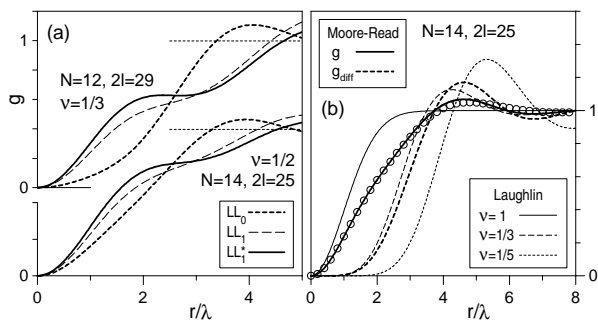}}
\caption{
   (a) Pair-distribution functions $g(r)$ of lowest $L=0$ states of 
   finite systems corresponding to $\nu={1\over3}$ and ${1\over2}$, 
   for pseudopotentials of electrons in the first and second LL, and 
   of CF's in the second LL.
   (b) The total $g(r)$ and ``remainder'' $g_{\rm diff}(r)$ curves 
   of the Moore--Read $\nu={1\over2}$ state; circles mark a fitting 
   linear combination of the curves for Laughlin states.}
\label{fig5}
\end{figure}

The evolution of $g(r)$ when going from the lowest electron LL to the 
second CF LL (i.e., from LL$_0$ to CF-LL$_1$) is clear when using a model 
pseudopotential $V_\zeta(\mathcal{R})=\zeta\,\delta_{\mathcal{R},1}+
(1-\zeta)\,\delta_{\mathcal{R},3}$.
For $\zeta\approx0$ or 1, the correlations (avoidance of $\mathcal{R}
=1$ or 3) are insensitive to $\zeta$, and both Laughlin and QE--QE
correlations are accurately reproduced by $V_0$ and $V_1$, respectively.
Modeling correlations among electrons in LL$_1$ (the second LL) is 
more difficult, because they are very sensitive to the exact form of
$V(\mathcal{R})$ at the corresponding $\zeta\sim{1\over2}$.
As a result, the $N$-electron Coulomb eigenstates in LL$_1$ are more
susceptible to finite-size errors than in LL$_0$ or CF-LL$_1$.
In large systems, a good trial state is only known at $\nu={1\over2}$
(Moore--Read state), and much less is established about the correlations 
at $\nu={1\over3}$.
Still, the $g(r)$ curves for electrons in LL$_1$ must certainly fall 
between the two extreme curves for $\zeta=0$ and 1 (and differ from 
both of them).
This is shown in Fig.~\ref{fig5}(a) for both $\nu={1\over3}$ and 
${1\over2}$.

The exact Moore--Read wavefunctions were calculated on a sphere for 
$N\le14$ and $2l=2N-3=25$ by diagonalizing a short-range three-body 
repulsion.\cite{3body}
In Fig.~\ref{fig5}(a) we only plotted $g(r)$ for $N=14$ because the 
$N=12$ curve is too close to be easily distinguished.
The values of $\alpha=0.602$ and 0.600 for $N=12$ and 14.
The $g_{\rm diff}(r)$, also shown, is positive definite, very 
different from the QE curves in Fig.~\ref{fig4}(b), and rather close 
to $g_1(r)$, where $g_p$ describes a Laughlin $\nu=(2p+1)^{-1}$ 
state.
Assuming $\alpha_{\rm MR}={3\over5}$ and expanding $g_{\rm diff}$ 
into $g_1$ and $g_2$ in accordance with Eq.~(\ref{eq2}) one
obtains an approximate formula
\begin{equation}
   g_{\rm MR}(r)\approx{3\over5}\,g_0(r)+{3\over10}\,g_1(r)
                                        +{1\over10}\,g_2(r),
\end{equation}
marked with the circles in Fig.~\ref{fig4}(b) that appears to be quite 
accurate (the largest finite-size error is in $g_2$ calculated for 
only $N=8$, while $g_1$ is for $N=12$).

The fact that $g_{\rm diff}$ is positive and rather featureless 
(similar to $g_p$) for the Moore--Read wavefunction is in contrast
with the result for QE's.
This difference may indicate that the QE clusters cannot be understood 
as literally as Moore--Read pairs.
Indeed, even the lack of correlation between the occurrence of $L=0$ 
ground states (or size of the excitation gap) and the divisibility 
of $N$ by $K=2$ or 3 precludes such a simple picture.
The fact that $g_{\rm diff}(r\sim3\lambda)<0$ could mean that the 
average relative (with respect to center of mass) angular momentum 
$\mathcal{R}_K$ of the QE clusters is much larger than 
$\mathcal{R}_K^{\rm min}={1\over2}K(K-1)$.
Certainly, $\mathcal{R}_K$ is only conserved for an isolated cluster, 
but it is possible that the QE clusters are more relaxed due to 
cluster--cluster interaction than the Moore--Read pairs are.
This would make $g_0$ underestimate the radius of the actual 
intra-cluster QE-QE correlation hole, and explain the negative sign 
of $g_{\rm diff}$.

\section{Conclusion}

From exact numerical diagonalization on Haldane sphere, we obtained
the energy spectra and wavefunctions of up to $N=14$ interacting 
Laughlin QE's (CF's in the second LL).
We identified the series of finite-size liquid ground states with 
a gap, which extrapolate to the experimentally observed incompressible 
FQH states at $\nu_e={4\over11}$, ${3\over8}$, and ${5\over13}$.
In these states, we calculated QE--QE pair-distribution functions 
$g(r)$, and showed that they increase as $\sim r^2$ at short range 
and have a pronounced shoulder at a medium range.
This behavior supports the idea of QE cluster formation, suggested
earlier from the analysis of QE--QE interaction pseudopotential.
The $g(r)$ is decomposed into short- and long-range contributions,
interpreted as correlations between the QE's from the same or different 
clusters.
The intra-cluster contribution to $g(r)$ is that of a full LL, and the 
remaining term identified with the inter-cluster QE--QE correlations 
appears to be the same in all three $\nu={1\over3}$, ${1\over2}$, and 
${2\over3}$ states.
The (average) cluster size on the other hand does depend on $\nu$, 
and we present arguments which suggest that the QE's form pairs at 
$\nu={1\over3}$ and triplets at $\nu={1\over2}$.

The qualitative difference between the $g(r)$ curves obtained here
for correlated CF's and those known for the Laughlin and Moore--Read 
liquids of electrons are another indication that the origin of 
incompressibility at $\nu_e={4\over11}$, ${3\over8}$, and ${5\over13}$
is different.
Of other hypotheses invoked in literature and mentioned here in the 
introduction, the broken-symmetry states cannot be excluded by our 
calculation in spherical geometry.
However, we anticipate that the QE's form a liquid (studied in this 
paper) also in experimental samples, because of the whole series of 
isotropic ground states with a gap occurring in finite systems of 
different size.

The authors thank W.~Pan, W.~Bardyszewski, and L.~Bryja for helpful 
discussions.
This work was supported by Grant DE-FG 02-97ER45657 of the Materials
Science Program -- Basic Energy Sciences of the U.S. Dept.\ of Energy 
and Grant 2P03B02424 of the Polish KBN.

\end{document}